# Social capital may mediate the relationship between social distance and COVID-19 prevalence

Keisuke Kokubun[1*]

1 Economic Research Institute, Japan Society for the Promotion of Machine Industry

## Abstract

The threat of the new coronavirus (COVID-19) is increasing. Regarding the difference in the infection rate observed in each region, in addition to studies seeking the cause due to differences in the social distance (population density), there is an increasing trend toward studies seeking the cause due to differences in social capital. However, studies have not yet been conducted on whether social capital could influence the infection rate even if it controls the effect of population density. Therefore, in this paper, we analyzed the relationship between infection rate, population density, and social capital using statistical data for each prefecture. Statistical analysis showed that social capital not only correlates with infection rates and population densities but still has a negative correlation with infection rates controlling for the effects of population density. Besides, controlling the relationship between variables by mean age showed that social capital had a greater correlation with infection rate than population density. In other words, social capital mediates the correlation between population density and infection rates. This means that social distance alone is not enough to deter coronavirus infection, and social capital needs to be recharged.

## Introduction

The threat of the new coronavirus (COVID-19), which was discovered in Wuhan, China in December 2019 and spread worldwide, is affecting the lives of many people without stopping. Japan is no exception, although it is often reported that its virus death rate is mysteriously low (Wingfield-Hayes, 2020). The number of infected people, which had subsided in June 2020, started increasing again in July. Hirata et al. (2020) showed that the infection rate tends to be high in large cities such as Tokyo and that the reason for this is the high population density of these large cities, and then stated that population density can be used as a proxy variable for social distance.

However, coronal infections cannot be prevented by social distances alone. In fact, in addition to securing social distance by refraining from going out, time lag, remote work, etc., the government is implementing anti-quarantine actions such as wearing a mask, encouraging hand and finger disinfection, and refraining from loud conversations as a basic policy for countermeasures against coronavirus infectious diseases, which has been strongly sought by the public (Cabinet Secretariat, 2020). Among some possible options, many researchers are paying attention to social capital as a factor that promotes these quarantine actions.

Social capital is the sharing of values, acceptance of norms, unity, and trust through reciprocity, and is said to play an important role in solving problems through interaction and cooperation (Coleman, 1988; Fukuyama, 1995; Putnam, 2000). Previous research has also shown that social capital is effective in solving public health challenges (Asri & Wiliyanarti, 2017; House, Landis, & Umberson, 1988; Pretty, 2003). For example, some studies show that the high level of social capital (trust and human connection to the government, reciprocity, reciprocity, and solidarity) influenced the practice of preventive actions such as vaccination, washing hands, and wearing a mask during the H1N1 outbreak in 2009 (Chuang et al., 2015; Ronnerstrand, 2013; See also the review by Pitas & Ehmer, 2020). In connection with the coronal disaster this time, studies that analyzed GPS information in the United States showed that residents in counties with high social capital were more cooperative in going out regulations (Borgonovi & Andrieu, 2020) and reduced the increase of coronavirus infection (Varshney & Socher, 2020). A study in Italy also showed that areas with high social capital tended to have low outing rates and low corona mortality (Bartscher et al., 2020). In a related study, research in the field of human resource management based on a questionnaire survey shows that social capital (reciprocal support between company support and employee organizational commitment) enhanced employees' willingness to cooperate in the workplace corona countermeasures (Kokubun, Ino, & Ishimura, 2020).

Why does social capital have this effect? This is because social capital encourages information sharing and removes the uncertainty of choice (Chung, Nam, & Koo, 2016; Li, Ye, & Sheu, 2014; Six et al., 2015), also encourages collective action and deters the occurrence of the free-rider problem (Koutsou, Partalidou, & Ragkos, 2014; Mosse, 2006; Ostrom, 2000; Pretty, 2003). Regarding the former, Kawachi & Berkman (2000) argue that social capital helps promote the dissemination of health-related information and prevents unhealthy conduct. Regarding the latter, an empirical study conducted in China after COVID-19 showed that social capital was successful in the prevention of epidemics such as refraining from going out and wearing masks (Yang &

Ren, 2020).

The latter, that is, the mechanism by which social capital restrains free riders, is also supported by game theory research. Generally speaking, for people to take coordinated actions in the relationships with others continuously, it is necessary to expect that long-term benefits of collaborating will surpass the short-term benefits of opportunistic actions, Therefore, to promote cooperative actions by increasing the benefits of cooperation and lowering the benefits of opportunism, it is necessary to have a mechanism in which a person who refuses cooperation receives "tippling" from another person (Axelrod, 1984). For reference, Japan's *murahachibu* (making people who do not follow the rules of the village out of line) is wisdom based on such a mechanism.

As mentioned above, more and more studies tend to show that social capital leads not only to epidemics prevention up to now but also to the promotion of quarantine action and deterrence of infection in this time of Corona. However, to the best of my knowledge, no studies have shown that even if the effects of population density (social distance) are removed, there is a negative correlation between high social capital and coronavirus infection rates. Therefore, in this study, I would like to address this issue using cross-sectional data at the local government level in Japan.

## Review of previous studies and presentation of hypotheses
**Population density and coronavirus infection rate**
The relationship between population density and infection rate is shown in Hirata et al. (2020). A study in China also confirmed a positive correlation between population density and infection rates before the lockdown (Wang et al., 2020). These researchers point out increased opportunities for human contact (smaller social distances) associated with high population densities as a reason for the correlation. In support, some studies have directly investigated the effects of social distance. For instance, a previous study based on the results of an independent online survey showed that the number of infected people from January to March this year tended to be smaller in prefectures with less face-to-face communication, use of public transportation, and fewer meals outsides (Shoji et al., 2020). Therefore, the following hypothesis is derived.
H1: There is a positive correlation between population density and infection rate.

**Social capital and coronavirus infection rates**
Previous studies have shown that social capital helps promote public health (Asri, Nuntaboot, and Wiliyanarti, 2017; Chuang et al., 2015; House et al. 1988; Pitas & Ehmer, 2020; Ronnerstrand, 2013; Pretty 2003). Besides, more and more studies at this time of Corona Eruption show that social capital encourages cooperation against coronavirus measures (Kokubun, Ino, & Ishimura, 2020; Yang & Ren, 2020) and control infections. (Bartscher et al., 2020; Varshney & Socher, 2020). Therefore, the following hypothesis is derived.

H2: There is a negative correlation between social capital and the infection rate.

**Population density and social capital**

Small cities are more likely to form social links than large cities (Putnam, 1995). According to one theory, the basic services provided by the government in large cities are not provided in small cities with a low population density, which means that the unity and turnaround of residents to maintain alternative services will be required (Browne, 2001). In support, several studies have found a negative effect of population density on social interaction (Brueckner & Largey, 2008; Dempsey et al., 2011). Therefore, many studies have shown that densely populated and large cities have lower social capital (Andrews, 2009; Eriksson & Rataj, 2019; Rupasingha et al., 2006; Veenstra, 2002). Therefore, the following hypothesis is derived.

H3: There is a negative correlation between population density and social capital.

**Partial mediation effect of social capital**

Furthermore, we would like to examine whether social capital correlates with the infection rate even if population density is included as a variable. The result of one empirical study recently conducted in Japan shows that the higher the population density, the greater the impact of trust with the surrounding people on health (Sato et al., 2018). This suggests that population density alone is not enough to predict people's health and that social capital acts as an intermediary between population density and health. Further, given that population density is not a variable that captures all social distances, and that social capital is involved in the prevention of various types of epidemics, as shown above, population density and social capital are thought to affect the infection rate complementing each other. Therefore, the following hypothesis will be established.

H4: Social capital partially mediates the relationship between population density and infection rates.

**Full mediation effect of social capital**

The population density used in this paper is merely an indicator of one aspect of social distance, in which the population is concentrated in a certain area. Even in densely populated areas, it is not impossible for people's efforts to reduce contact with each other.

For example, a US study using mobile GPS location data showed that the normative element of social capital correlated with social distance measured by the length of time at home (Bai, Jin, & Wan, 2020). Further, infection risk will depend to a large extent on each person's social capital following norms and disciplines such as remote work, hand washing, and disinfection. Therefore, the following hypothesis is derived.

H5: Social capital fully mediates the relationship between population density and infection rates (Social capital has a greater correlation with infection rate than population density).

## Data

For the coronavirus infection rate, the cumulative number of infected people per 1 million population by July 3, 2020, calculated by Sapporo Medical University (2020), was used. For the population density, the "population density per 1 $km^2$ of habitable area" for 2018 recorded in the Statistics Bureau (2020) was used. Both the infection rate and population density were log-transformed and used for the analysis. For social capital, we used the "social capital comprehensive index" recorded in the Cabinet Office (2003). This index is calculated from the response of 3,878 people to a questionnaire survey consisting of 10 questions about "friendship/interaction", "trust" and "social participation", combining the figures in other two comprehensive statistics about "volunteer activity participation rate" and "community donation amount per person". It is standardized by the formula "(prefecture value-average value)/standard deviation". This index is used because it is one of the most comprehensive measures of social capital in Japan, and it consists of both subjective information (questionnaire survey) and objective information (external source), increasing objectivity and reducing the common method bias. Also, as a strength not found on other similar scales, empirical studies have shown that this index correlates with lower unemployment and crime rates, higher fertility rates, higher life expectancy, and higher new start-up rates (Cabinet Office, 2003). Also, the average age for 2015 was obtained from the Statistics Bureau (2017) for use as a control variable.

## Analysis and results

Table 1 is descriptive statistics. The lower part of the correlation coefficient table is the normal correlation coefficient, and the upper part is the partial correlation coefficient controlled by the average age. In both cases, significant correlations are shown among the three variables. Figures 1 to 3 are scatter diagrams showing these relationships. These support H1 to H3.

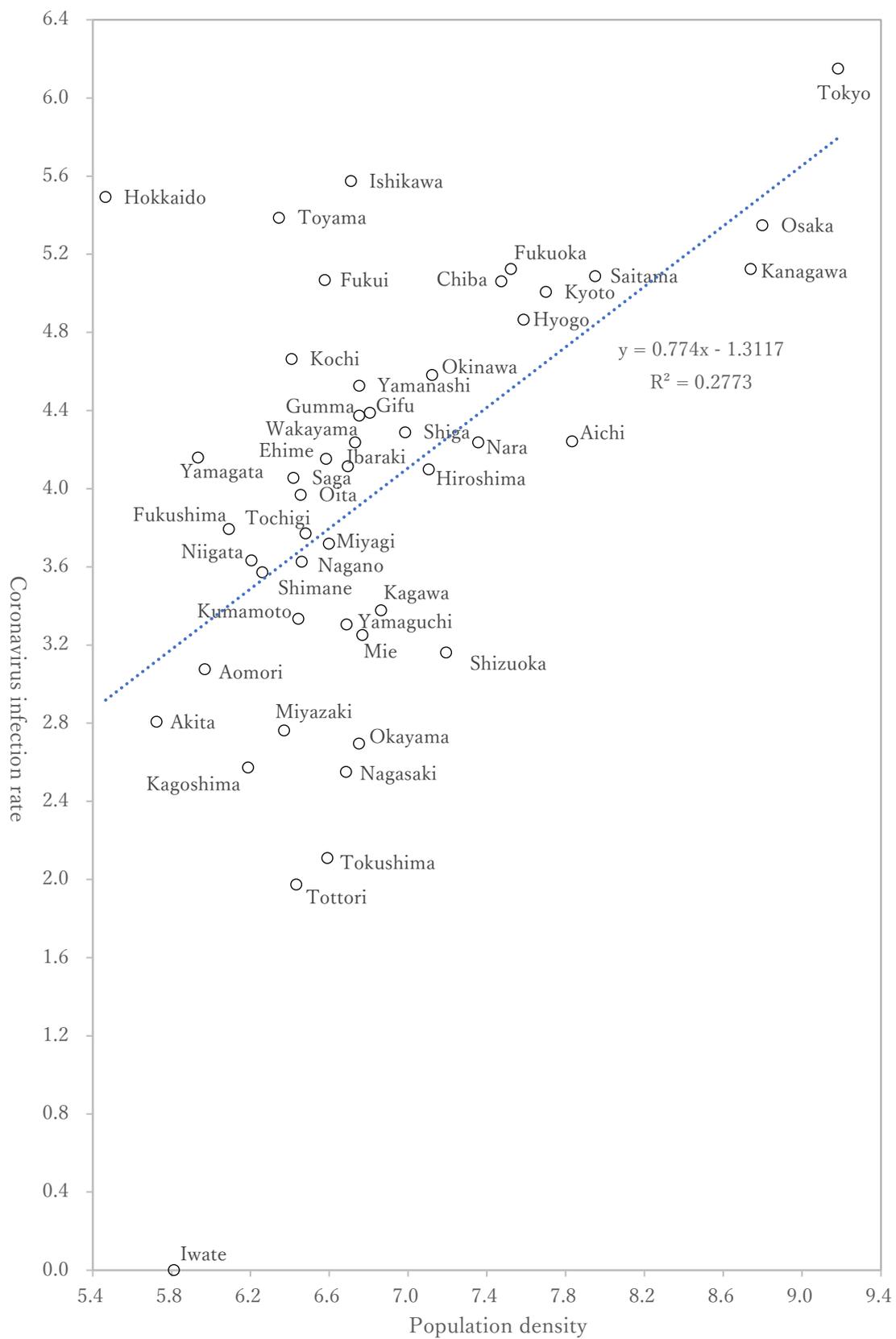

Figure 1. Correlation between population density and coronavirus infection rate

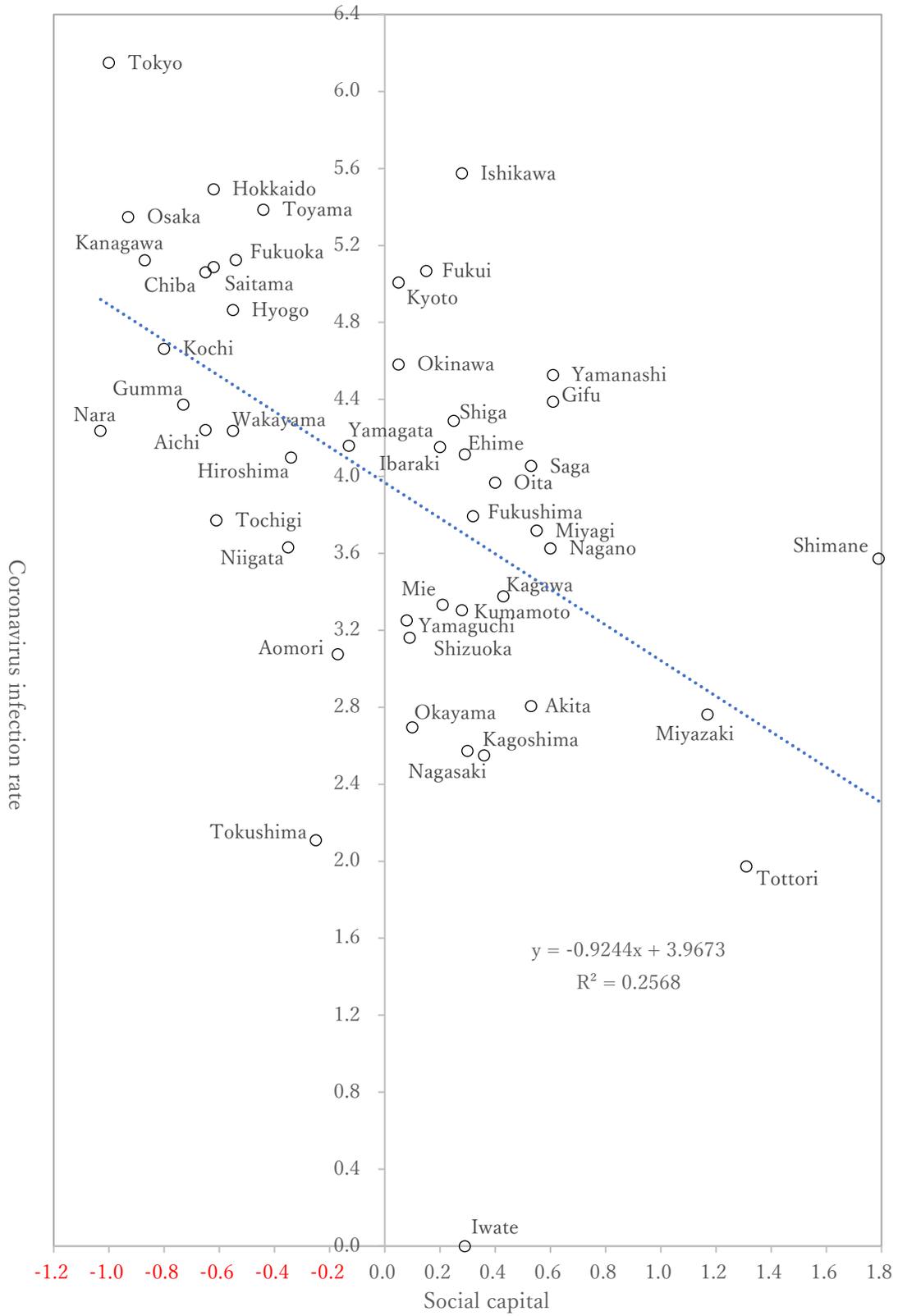

Figure 2. Correlation between social capital and coronavirus infection rates

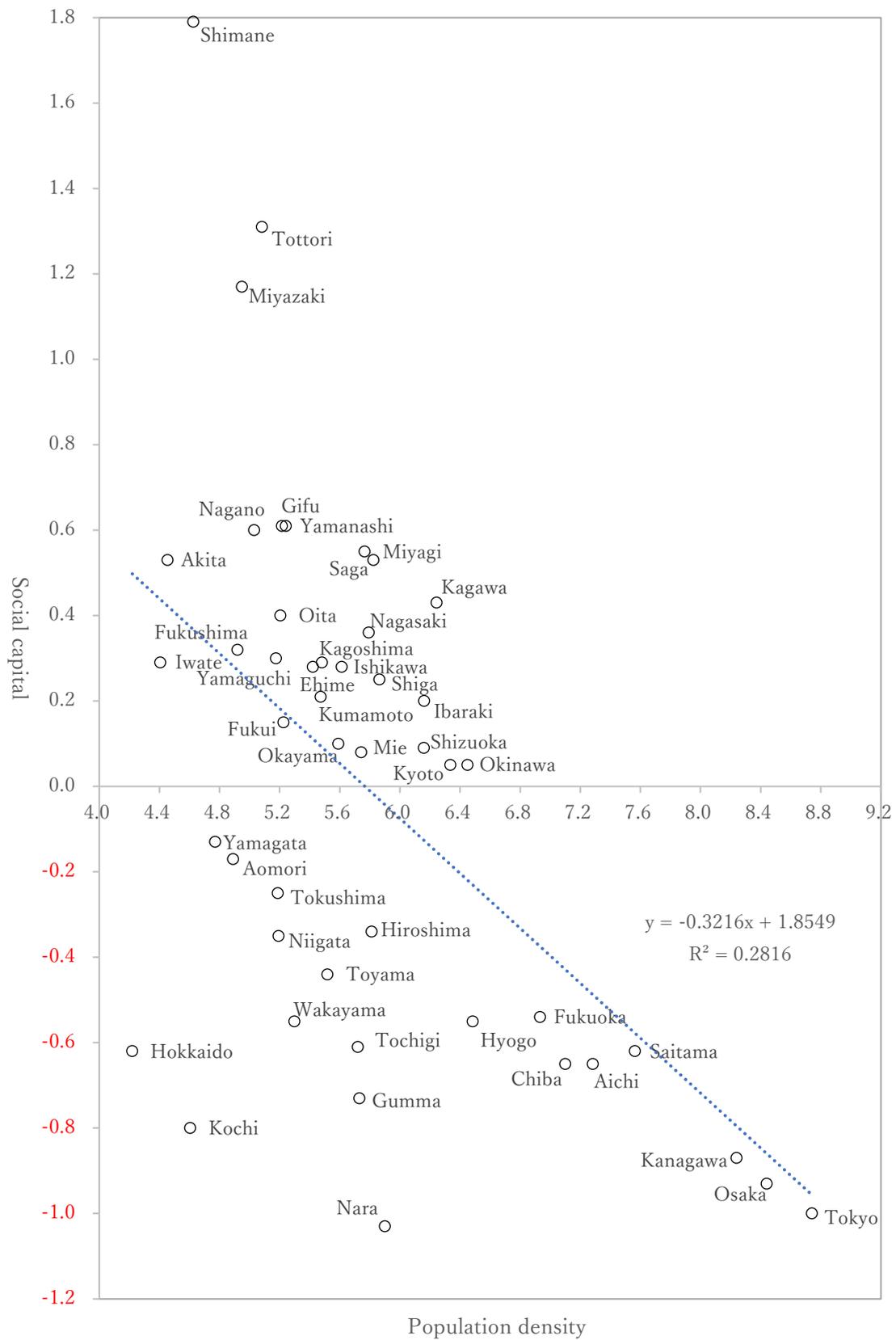

Figure 3. Correlation between population density and social capital

Table 1. Descriptive statistics

|   |   | Mean | SD | 1 | 2 | 3 |
|---|---|---|---|---|---|---|
| 1 | Coronavirus infection rate | 3.967 | 1.132 |  | 0.310* | -0.416** |
| 2 | Population density | 6.821 | 0.770 | 0.527** |  | -0.398** |
| 3 | Social capital | 0.000 | 0.621 | -0.507** | -0.506** |  |
| 4 | Average age | 47.300 | 1.667 | -0.475** | -0.697** | 0.342* |

Note: ** Significant at 1% level.   * Significant at 5% level.   n = 47

Coronavirus infection rate: Number of infected people per million (logarithmic display). For Iwate prefecture where the number of infected people was 0 (that means logarithmic conversion not possible), we substituted 0 assuming that there was one infected person there. Population density: Population density per 1km$^2$ of habitable area (logarithmic display). Social capital: A standardized index consisting of "communications and exchanges", "trust", "social participation", "volunteer activity rate", and "community solicitation amount per capita".

Furthermore, the results of the path analysis shown in Fig. 4 indicate that the correlation between social capital and infection rate remains even when population density is controlled, that is, social capital partially mediates the relationship between population density and infection rate. This supports H4. Furthermore, Figure 5 shows the results of a model including age as a control. Here, population density makes the path of infection rates insignificant, indicating that social capital mediates the relationship between population density and infection rates completely. This indicates that social capital influences the infection rate more than population density, and it can be said to support H5.

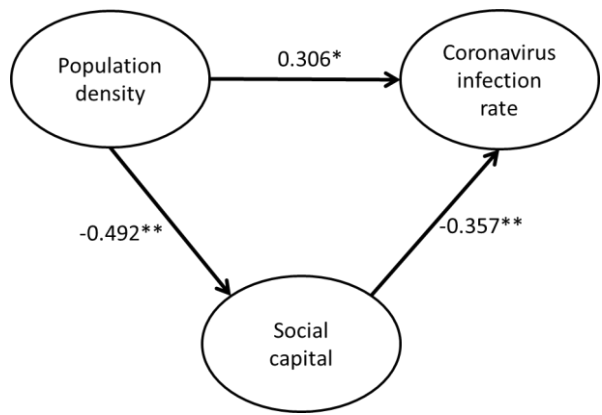

Figure 4. Result of path analysis
Note: ** Significant at 1% level.   * Significant at 5% level.   n = 47

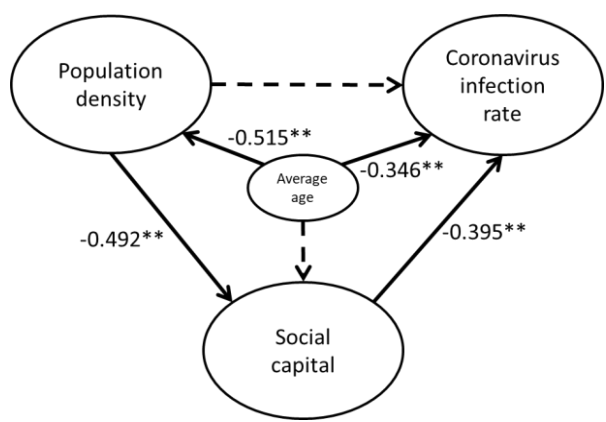

Figure 5. Result of path analysis (Controlled by average age)
Note: ** Significant at 1% level.   * Significant at 5% level.   n = 47
Dashed arrow indicates not significant at 5% level.
Goodness-of-fit indices: $\chi2$ = 2.150; df = 2; root mean square error of approximation (RMSEA) = 0.040; probability of close fit (PCLOSE) = 0.382; goodness of fit index (GFI) = 0.977; adjusted goodness of fit index (AGFI) = 0.884; normed fit index (NFI) = 0.957; comparative fit index (CFI) = 0.997.

**Discussion**

The results in this paper show that the negative correlation between social capital and infection rates is still statistically significant in controlling population density. It can be said that social capital partially mediates the relationship between population density and infection rates. To put it another way, securing social distance is not enough to prevent infection, and it is essential to foster social capital based on solidarity, trust, and adherence to norms. Also, age-controlled models show that social capital influences infection rates more than population density. As shown in Figures 1-2, Hokkaido is a good example of this. Low population density should lead to low infection rates but shows high infection rates due to low social capital. This can be said to be the result of supporting the argument by Kitayama et al. (2006), which states that the culture of the pioneering spirit inherited in Hokkaido is based on individualism and empirical research by Yamawaki (2012) who verified it.

    Looking back on this time's coronal disaster based on the results of this article, the high infection rate in large cities such as Tokyo can be explained not only by the narrow social distance but also by the weakness of social capital. While the dangers of corona are being exclaimed, the reason why they go out to the so-called *yoru no machi* (entertainment district) where the risk of infection is high is that the norms for limiting their behaviors are weak. If this is a local city where a culture of traitor-tattooing is alive,

there will be an immediate rumor of "What is that person thinking?", which can be a strong weapon that prevents his/her imprudent behavior. Especially in situations where the experience values cannot be utilized like in the case of the Corona Era this time, it is considered that the bonds and norms between citizens played a major role in controlling people and taking appropriate actions, especially in local cities.

If so, from the perspective of preventing infectious diseases, it will be necessary to take various measures to restore social capital in each prefecture, especially in the heavily damaged large cities. In other words, instead of pursuing only the fulfillment of one's desires, it is necessary to develop a culture where people respect cooperation and norms and are willing to control their actions for the people around them. However, it is important to note that these cultures can sometimes have negative consequences, such as the exclusion of strangers (Portes, 1998). Furthermore, it has often been pointed out that such exclusivity hinders the internationalization of Japanese organizations (Keeley, 2001) especially in acquiring and motivating high-class or diverse human resources (Kokubun, 2018; Kokubun and Yasui, 2020). Everything has its pros and cons. It may also be said that the reason why the prevention of epidemics in Japan, especially in local cities, looks more successful than some other countries so far is that the good aspects of the Japanese as an *inakamono* (rural person) works. Cultivating a spirit that is open to others while at the same time strengthening unity is the right way to manage crises in a global society. The national, local governments, schools, and workplaces must work in tandem to achieve this goal.

**Implication**

In this paper, we have shown by cross-section analysis using prefecture-level data in Japan that social capital plays an important role in addition to the social distance in preventing infection in this coronal epidemic. This finding can be utilized not only for the measures for the second wave of coronavirus expected shortly but also for the preventive measures for various infectious disease viruses in the future.

**Limitation**

In this paper, we used population density as a proxy variable for social distance, following Hirata et al. (2020). However, in the real world, a kind of social distance that cannot be supplemented by population density is practiced, such as avoiding congestion due to staggered hours. Further, the magnitude of these efforts has been shown in previous studies to be influenced by social capital. Therefore, the difference in the definition of social distance does not seem to significantly change the main conclusion of this paper.

Rather, the larger limit would be the small size of the sample. If the infection rate, social capital, and social distance data with smaller units can be obtained from local governments, it is considered significant to verify the reproducibility of the results in this paper in future research.

## Conclusion

The threat of the new coronavirus (COVID-19) is increasing. Regarding the difference in the infection rate observed in each region, in addition to studies seeking the cause due to differences in the social distance (population density), there is an increasing trend toward studies seeking the cause due to differences in social capital. However, studies have not yet been conducted on whether social capital could influence the infection rate even if it controls the effect of population density. Therefore, in this paper, we analyzed the relationship between infection rate, population density, and social capital using statistical data for each prefecture. Statistical analysis showed that social capital not only correlates with infection rates and population densities but still has a negative correlation with infection rates controlling for the effects of population density. Besides, controlling the relationship between variables by mean age showed that social capital had a greater correlation with infection rate than population density. In other words, social capital mediates the correlation between population density and infection rates. This means that social distance alone is not enough to deter coronavirus infection, and social capital needs to be recharged.

Pitas, N., & Ehmer, C. (2020). Social capital in the response to COVID-19. *American Journal of Health Promotion*, 0890117120924531. https://doi.org/10.1177/0890117120924531

Portes, A. (1998). Social capital: Its origins and applications in modern sociology. *Annual Review of Sociology, 24*(1), 1-24. https://doi.org/10.1146/annurev.soc.24.1.1

Pretty, J. (2003). Social capital and the collective management of resources. *Science, 302*(5652), 1912-1914. https://doi.org/10.1126/science.1090847

Putnam, R. D. (1995). Tuning in, tuning out: the strange disappearance of social capital in America. *PS: Political Science & Politics, 28*(4), 664–683.

Putnam, R. D. (2000). *Bowling alone: The collapse and revival of American community*. New York, NY: Simon & Schuster.

Ronnerstrand, B. (2013). Social capital and immunization against the 2009 A(H1N1)V pandemic in Sweden. *Scandinavian Journal of Public Health, 41*(8), 853–859. https://doi.org/10.1016/j.puhe.2014.05.015

Rupasingha, A., Goetz, S. J., & Freshwater, D. (2006). The production of social capital in US counties. *The Journal of Socio-Economics, 35*(1), 83-101. https://doi.org/10.1016/j.socec.2005.11.001

Sapporo Medical University (2020). *Todohukenbetsu Jinkoatarino Shingata coronavirus Kansenshasuno Suii(Changes in the number of new coronavirus infections per population by prefecture)*. Department of Genome Medical Science, Frontier Medical Research Institute, Sapporo Medical University. https://web.sapmed.ac.jp/canmol/coronavirus/japan.html? (accessed: 2020.7.6)

Sato, Y., Aida, J., Tsuboya, T., Shirai, K., Koyama, S., Matsuyama, Y., Kondo, K., & Osaka, K. (2018). Generalized and particularized trust for health between urban and rural residents in Japan: A cohort study from the JAGES project. *Social Science & Medicine, 202*, 43-53. https://doi.org/10.1016/j.socscimed.2018.02.015

Shoji, M., Cato, S., Iida, T., Ishida, K., Ito, A., & McElwain, K. (2020). COVID-19 and social distancing in the absence of legal enforcement: Survey evidence from Japan. *MPRA Paper 100723*.

Six, B., Zimmeren, E., Popa, F., & Frison, C. (2015). Trust and social capital in the design and evolution of institutions for collective action. *International Journal of Commons, 9*(1), 151–176. http://doi.org/10.18352/ijc.435

Statistics Bureau (2017). *Heisei 27nen Kokuzeichosa: Todofuken Shichosonbetsu Tokeihyo (2015 Census Statistics Table by Prefecture/City)*, Statistics Bureau, Ministry of Internal Affairs and Communications, Tokyo.

Statistics Bureau (2020). *Shakaiseikatsu tokeisihyo: Todofukenno sihyo, 2020 (Social Life*
15